# Concerning impact of the quantum vacuum on orbits of planets

Dragan Slavkov Hajdukovic
INFI, Cetinje, Montenegro
Email: dragan.hajdukovic@cern.ch

**Abstract**. We point out serious shortcomings of a very recent article (Iorio in Astrophys. Space Sci. 364:126, 2019) *wrongly claiming* that the current precision with which we know orbits of planets in the Solar System rules out the possibility of gravitational polarization of the quantum vacuum. The main mistake is that the Sun and a planet are considered as an isolated binary system completely neglecting the existence of other planets and their crucial contribution to the gravitational polarization of the quantum vacuum.

The claim of a very recent article (Iorio, 2019) is that the current precision with which we know the perihelion precession of orbits of planets in the Solar System rules out the possibility of gravitational polarization of the quantum vacuum (Hajdukovic 2011, 2014).

Before the presentation of very serious shortcomings of this article let us underscore that the theory presented in a series of articles (See for instance: Hajdukovic 2011, 2013, 2014) is not a "modified gravitational theory" as wrongly understood and interpreted in Iorio (2019). Our theory is not a modification of Newton's law of gravity. In fact, the quantum vacuum is considered as a so far "forgotten" (i.e. neglected) source of gravity, but the *effective* gravitational charge (gravitational mass) of the quantum vacuum acts just as any ordinary mass.

In the Newtonian theory of gravity, perihelion precession exists *only if* there is a departure from spherical symmetry. Consequently, the perihelion precession is zero in the case of an ideal, isolated binary, composed of two point-like bodies orbiting about their common centre of mass; on the other hand, in a real binary (but still assuming isolation), a usually small precession exists because bodies in the binary are not perfect spheres. Of course, our Solar System is not a binary but a multibody system, and, the crucial point is that *the dominant part of the perihelion precession of a planet is caused by other planets*. For instance (See Table 3 in Park et al. 2017), Venus and Jupiter contribute respectively about 48.2% and 26.8% to the total precession of Mercury, while Mars contributes 0.42%. If we neglect the tiny contribution of Mars, there will be disagreement between theoretical prediction and the measured value, while if we neglect all planets the result would be meaningless. In brief, if the impact of other planets is neglected, the calculated Newtonian perihelion precession of a planet is wrong, i.e. incompatible with the observed precession.

This well-established impact of other planets makes questionable the approach (Iorio, 2019) in which the gravitational polarisation caused by other planets is completely neglected. Let us make much clearer the importance of the complexity of distribution of the *effective* gravitational charge of the quantum vacuum.

The starting point is that the quantum vacuum can be considered as a source of gravity thanks to the hypothesis that *quantum vacuum fluctuations* (or in more popular wording particle-antiparticle pairs) are *virtual gravitational dipoles* (See for instance: Hajdukovic 2011, 2014, 2019). A classical gravitational dipole is defined in full analogy with the well-known electric dipole; hence a gravitational dipole is a system of two particles that have the same inertial mass but the opposite gravitational charge (gravitational mass).

An external gravitational field can align gravitational dipoles in the same way as an external electric field can align electric dipoles. Hence, the gravitational polarization density $\mathbf{P}_g$, i.e. the gravitational dipole moment per unit volume, can be attributed to the quantum vacuum. It is obvious that the magnitude of the gravitational polarization density $\mathbf{P}_g$ satisfies the inequality $0 \leq |\mathbf{P}_g| \leq P_{gmax}$, where 0 corresponds to the *random orientations* of dipoles, while the maximal magnitude $P_{gmax}$ corresponds to the case of *saturation* (when all dipoles are aligned with the external gravitational field).

The main result is that the quantum vacuum can be considered as an omni-present fluid with an *effective gravitational charge density*:

$$\rho_{qv} = -\nabla \cdot \mathbf{P}_g . \tag{1}$$

If planets and other celestial bodies are neglected, according to Eq. (1), there is *a single halo* of the polarized quantum vacuum around the Sun and it is the innermost part of the region of saturation. Within *a single halo model*, the only gravitational effect of the quantum vacuum is a tiny constant acceleration towards the Sun. This additional acceleration (caused by the quantum vacuum and not by modification of gravity) is estimated to be $g_{qvmax} = 4\pi G P_{gmax} \approx 5 \times 10^{-11}$ m/s$^2$ ; note that this acceleration is about 20 times smaller than quite a different acceleration in the Pioneer anomaly. A major impact of this acceleration, on a planetary orbit (with semimajor axes *a* and eccentricity *e*) is a *retrograde* perihelion precession per orbit:

$$\Delta\omega_{qvSun} = -2\pi\sqrt{1-e^2}\frac{a^2}{GM_{Sun}}g_{qvmax}. \tag{2}$$

The word "Sun" is used in Eq. (2) in order to avoid misunderstandings and make clear that the corresponding *retrograde precession* is caused *exclusively* by the halo of the Sun. The numerical values of this "anomalous" precession (see Table 1 in Hajdukovic, 2013) are much smaller than the corresponding Newtonian precession.

However, this simple picture of a single halo is not correct, because all planets and smaller celestial bodies have their own halos of the polarised quantum vacuum. For instance, near the Earth, the gravitational field of the Earth is much stronger than the gravitational field of the Sun and other bodies; hence dipoles are oriented towards the Earth, not towards the Sun, and consequently the Earth has its own halo marked by saturation. Of course, corresponding to each halo is an effective gravitational charge; consequently, masses of all celestial bodies are slightly increased by their individual halos. In addition to dipoles that point to a body there are also dipoles aligned with a resultant gravitational field that doesn't point to any Solar System body. Hence, a single halo model is "blind" to two major (for clarity we will call them "non-solar") impacts of the quantum vacuum: *modification of the mass of bodies by their individual halos* and the *impact of dipoles which do not belong to any individual halo*. Unfortunately, the "non-solar" contribution of the quantum vacuum to perihelion precession cannot be calculated analytically; the use of numerical methods is inevitable. At this point it is crucial to underscore that the retrograde precession determined by Eq. (2) and the expected prograde "non-solar" precession (that remains to be determined numerically) would cancel to some unknown extent. In brief, the eventual perihelion precession of planets caused by the quantum vacuum is not known (the appropriate numerical calculations are a challenging task); consequently, a comparison with empirical evidence is not possible.

It is fair to say that there is a common "responsibility" for the astonishing mistake (Iorio 2019) to consider the Sun and a planet as an isolated binary and to compare a still unknown quantity with observations! The initial article (Hajdukovic 2013) was devoted to orbits of tiny satellites in trans-Neptunian binaries; apparently, some of these binaries are the most promising systems (Gai and Vecchiato 2014) to test the eventual gravitational impact of the quantum vacuum. Completely focused on trans-Neptunian binaries, the author was not sufficiently cautious to underscore the obvious fact that what is a good approximation for some (not all) trans-Neptunian binaries is not necessarily valid for a planet orbiting around the Sun. The rest of the "responsibility" may be attributed to the usually superficial approach of mainstream scientists towards radically new ideas.

The second shortcoming is a general one (i.e. not related to the existence of the quantum vacuum). Because of this second shortcoming, it is quite possible that even relatively big (and physically unlikely) precessions determined by Eq. (2) are not ruled out by the existing empirical evidence for planets. Let us explain this statement.

What astronomers measure with high accuracy is the *total* perihelion precession of a planet; *individual contributions are not measured but calculated* (for instance perihelion precession caused by Jupiter and precession caused by general relativistic effects are calculated, not measured). According to the position taken by astronomers (Iorio 2019 and references therein) each new source of precession is excluded if the predicted precession is larger than the uncertainty of the measurement; *the implicit assumption* is that the impact of all (so far included) sources of precession is accurately known, not leaving room for an additional precession (See the paper of Křížek 2017 that is relevant to this topic). This is questionable because in principle all calculations are model-dependent and recalculations within a slightly different model can produce tiny differences making possible the inclusion of an effect that is larger than the uncertainty of the measurement. Let us make this point of view clearer regarding solar system ephemerides.

As is well known, ephemerides are the result of a numerical integration of the dynamical equations of motion which describe the gravitational physics of the Solar System. Historically, the first ephemerides were created using only Newtonian gravity. Today, ephemerides include General Relativistic effects. What we propose is to include the quantum vacuum as well. More precisely, we propose to create new ephemerides, with the quantum vacuum, from the beginning included as a source of gravity, in the dynamical equations of motion. Comparison of "quantum vacuum" ephemerides with the existing ephemerides will reveal if the presumed gravitational impact of the quantum vacuum is compatible or not with the empirical evidence.

In conclusion the article of Iorio (2019) corresponds to what we have called above "a single halo" model, which is "blind" to the complexity of the impact of the quantum vacuum; consequently, the conclusions of the article are invalid. Hence, a crucial shortcoming of the mentioned study is the use of "a single halo" model, together with the existing ephemerides that neglect the quantum vacuum in the dynamical equations of motion; the complexity of quantum vacuum effects demands creation of new ephemerides, with the quantum vacuum included in equations of motion from the beginning. Hopefully astronomers will do it soon.